# Elastic, dielectric and piezoelectric anomalies and Raman spectroscopy of $0.5Ba(Ti_{0.8}Zr_{0.2})O_3$-$0.5(Ba_{0.7}Ca_{0.3})TiO_3$

Dragan Damjanovic[a][1], Alberto Biancoli[1], Leili Batooli[1], Amirhossein Vahabzadeh[1] and Joe Trodahl[2]

[1]*Ceramics Laboratory, Swiss Federal Institute of Technology in Lausanne - EPFL, 1015 Lausanne, Switzerland*

[2]*MacDiarmid Institute of Advanced Materials and Nanotechnology, Victoria University, P.O. Box 600, Wellington, New Zealand*

The solid solution $0.5Ba(Ti_{0.8}Zr_{0.2})O_3$-$0.5(Ba_{0.7}Ca_{0.3})TiO_3$ (BCZT) is a promising lead-free piezoelectric material with exceptionally high piezoelectric coefficients. The strong response is related to structural instabilities close to ambient temperature. We report here on temperature-induced anomalies in the dielectric, piezoelectric, and elastic coefficients and Raman spectroscopy of ceramic BCZT. The data indicate ferroelectric-ferroelectric structural phase transitions in this material in addition to those previously reported. An anomaly is also observed above the Curie temperature $T_C$ and is associated with the loss of polar structure that persists thirty degrees above $T_C$.

[a] Author to whom correspondence should be addressed. Electronic mail: dragan.damjanovic@epfl.ch.

Solid solutions $(1-x)Ba(Ti_{0.8}Zr_{0.2})O_3-x(Ba_{0.7}Ca_{0.3})TiO_3$ (or BCZT-100x/30/20, where the last two numbers correspond to atomic concentrations of Ca and Zr[1]) exhibit strong electro-mechanical response and are of potential interest as a lead-free alternative to $Pb(Zr,Ti)O_3$ (PZT). Of particular interest is composition with x=0.5 (BCZT-50/30/20) whose longitudinal piezoelectric coefficient $d_{33}$ is reported to be over 600 pC/N, a value higher than in soft PZT.[2] Liu and Ren[2] proposed that the phase diagram of BCZT-100x/30/20 is similar to that of PZT (see Fig. 1) in that it consists of three regions: paraelectric cubic (C) and two ferroelectric regions, rhombohedral (R) and tetragonal (T). In contrast to the nearly vertical morhotropic phase boundary separating R and T phases in PZT, this boundary is strongly temperature dependent in BCZT-100x/30/20.

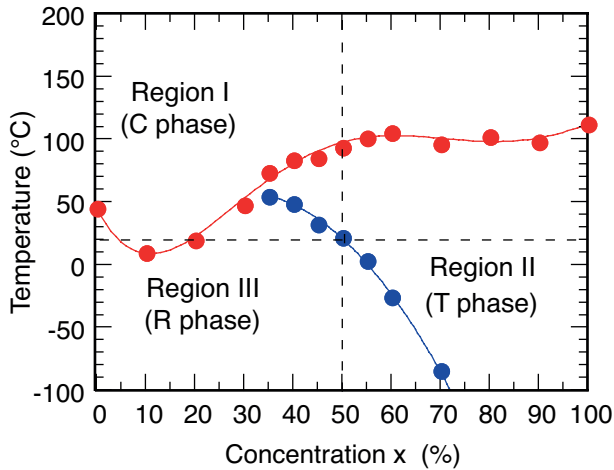

Figure 1 (color online). The phase diagram of BCZT-100x/30/20 as proposed in Ref. 2. The vertical dashed line indicates the position of BCZT-50/30/20 while the horizontal marks 20°C.

The strong responses have been interpreted in terms of isotropic flattening of a free energy profile associated with the C-R-T triple point which is also a tricritical point.[2] It is also clear from Fig. 1 that for BCZT-50/30/20 the R-T boundary lies close to the ambient temperature. In the proximity of such structural instabilities, ferroelectrics often exhibit large, anisotropic increase in piezoelectric, dielectric and elastic responses.[3] This is a well known result for single crystals of BaTiO$_3$, the parent system of BCZT.[4] Thus, it is not entirely surprising that anomalies in the electro-mechanical response, including a peak in d$_{33}$, were observed in BCZT-50/30/20 ceramics near the R-T boundary.[5] On heating from R-T toward C-T boundary the piezoelectric coefficient remains relatively large.[5] A possible reason for this is that R and T phases are separated by only about 70°C (Fig. 1); the enhanced response within T phase is thus driven by instabilities at both R-T and T-C transitions, each instability enhancing piezoelectric coefficients along different crystallographic directions.[5] The net result of two instabilities near each other is an isotropic increase of the coefficients. In this way the majority of grains in a ceramic sample exhibit enhanced response, regardless of their orientation with respect to the driving field. Thus, the narrow T phase resembles a triple point, missing only instability along the direct R-C path.[6]

The identification of the origin of large piezoelectric coefficients on the basis of phase instabilities only is not complete because as much as half of the piezoelectric response in ceramics may be associated with motion of domain walls.[7] However, important parameters of the domain wall contribution (e.g., domain structure, energy, mobility and width of domain walls) depend on crystal structure and have been discussed for BCZT-50/30/20 in Ref. 8. It is thus important for a detailed interpretation of the large piezoelectric response in BCZT to identify all crystal phases and domain-wall processes. The parent phase BaTiO$_3$ during cooling follows the C-T-Orthorhombic (O)-R sequence. Adding Zr to BaTiO$_3$ shifts the T-O and O-R boundaries toward higher temperatures while Ca has an opposite effect. In the solid solution these relations become more complex.[1,9-11] The phase diagram of BCZT-100x/30/20 proposed in Ref. 2 is based on dielectric data backed up by X-ray diffraction (XRD) and transmission electron microscopy (TEM) observations.[8] These studies suggest that tetragonal and rhombohedral phases of BCZT-50/30/20 are separated by a temperature region of unspecified width where the T and R phases coexist. No orthorhombic phase has been reported.

In addition to the T-R transition, the dielectric permittivity data for BCZT-50/30/20 in Figs. 1 of Refs. 2 and 5 suggest a weak hump and thermal hysteresis in the permittivity below room temperature. This anomaly was not discussed in Refs. 2 and 5 and could be caused by various processes in the material. We present here evidence, based on dielectric permittivity, piezoelectric resonance and elastic stiffness measurements, that BCZT-50/30/20 exhibits below room temperature an anomaly qualitatively comparable to the one occurring just above room temperature and identified in Refs. 2 and 8 as T-R transition. Another structural phase transition is indicated in the vicinity of -60°C by Raman spectroscopy.

BCZT-50/30/20 samples were prepared by mixing $BaZrO_3$ (98.5%; major metallic impurity Sr), $CaTiO_3$ (99.8%; major metallic impurity Ce) and $BaTiO_3$ (99.95 %) powders. The purities of powders are comparable to those used in Ref. 2. The mixed powders were calcined at 1150 °C for 4 hours, and after additional mixing for 8 hours and pressing into discs or rectangular bars they were sintered on a platinum foil at 1500 °C for 4 hours in air. The final density was 93% of the theoretical. XRD spectra, measured with a Bruker D8 Discover diffractometer at 20°C, indicate pseudocubic symmetry with splitting of {200} peaks similar to those reported in Ref. 2 i.e., resembling mixture of R and T phases. No secondary phases are apparent. For electrical measurements pellets were electroded with Cr-Au electrodes. For piezoelectric measurements samples were poled at 20 kV/cm applied for 30 minutes at 40°C. The capacitance and impedance needed for determination of permittivity and resonance parameters were measured with an HP 4284A LCR bridge and an HP 4194A Impedance Analyser. The mechanical stiffness (storage modulus and loss) was measured in the single cantilever mode with a Perkin-Elmer PYRIS Diamond Dynamic Mechanical Analyzer (DMA). All measurements were made during cooling and heating with the rate of 1-2 °C/min. Routine Raman spectra were taken with a Jobin-Yvon LabRam HR spectrometer, and for work to low frequencies with a T64000 triple monochromator. All spectra were collected using the 514 nm Ar+ line. The sample was placed in a Linkam variable-temperature microscope stage and the temperature was allowed to stabilize for at least two minutes before spectra were recorded.

The dielectric permittivity $\varepsilon'$, elastic storage modulus $E'$ and corresponding losses, $\varepsilon''$ and $E''$, are shown during heating of an unpoled sample in Fig. 2. The ferroelectric-paraelectric phase transition (T-C) is clearly visible around 90°C, as is the phase transition just above room temperature, identified in Ref. 2 as R-T. Another anomaly, which is barely discernable in $\varepsilon'$, is clearly seen below room temperature in $\varepsilon''$, $E'$ and $E''$. It is not unusual in ferroelectrics that a phase transition strongly affects elastic properties but is not mirrored in their dielectric response. In PZT, for example, onset of oxygen octahedra tilts leaves a strong signal in the temperature dependence of the elastic compliance and a very weak one in the dielectric permittivity.[12] Another example where elastic and dielectric transitions are decoupled has been reported in $Ba_2NaNb_5O_{15}$.[13]

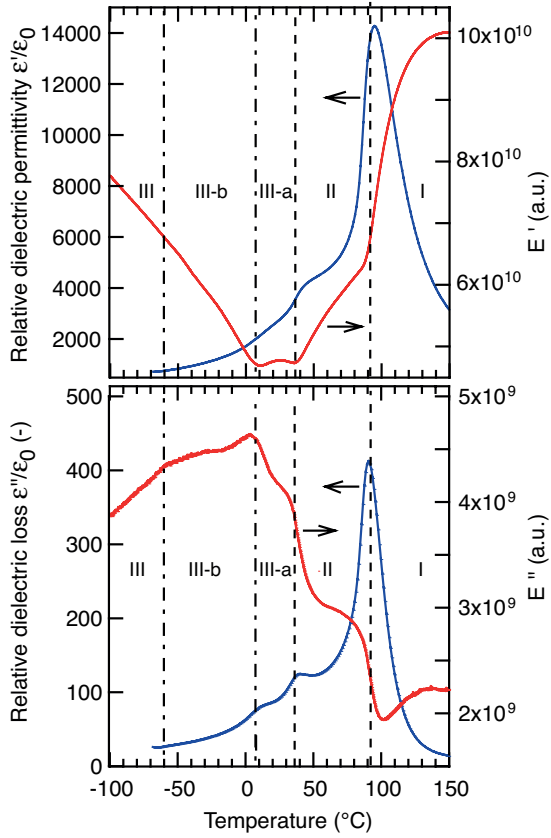

Fig. 2 (color online). (top) Relative dielectric permittivity $\varepsilon'/\varepsilon_0$ and storage modulus $E'$ and (bottom) mechanical and dielectric losses, $\varepsilon''/\varepsilon_0$ and $E''$, as a function of temperature for BCZT-50/30/20. Four anomalies are clearly distinguishable in $E'$, $E''$ and $\varepsilon''/\varepsilon_0$, indicating four distinct regions, I, II, III-a, III-b and III that are separated with the dashed and dot-dashed lines. The anomalies identified in this work are marked by the dot-dashed lines. The dielectric data were taken at 100 Hz and mechanical at 1 Hz. The elastic modulus is expressed in arbitrary units. Because BCZT cantilevers are fragile, they could not be completely clamped in the sample holder. Thus, numerical values of $E'$ and $E''$ shown in Figure 2a in Pa are somewhat underestimated and units are marked as "arbitrary". Compare with data in ref. 5.

Figure 3 shows that anomalies in $E'$ and $\varepsilon'$ exhibit thermal hysteresis, as would be expected for first order phase transitions. The elastic anomaly just below room temperature is also clearly visible in the elastic compliance determined from piezoelectric resonance data (not shown here). The significance of this result will be apparent later.

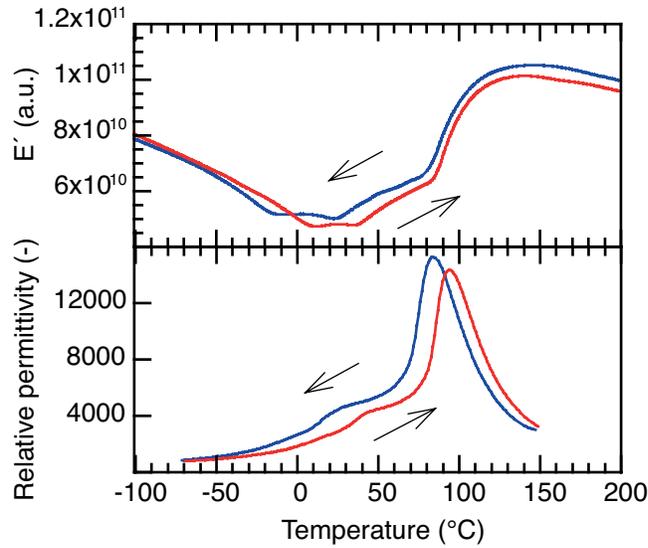

Fig. 3 (color online). Thermal hysteresis of (top) the elastic storage modulus and (bottom) relative dielectric permittivity of BCZT-50/30/20.

Raman spectroscopy has been undertaken to seek evidence of structural phase transitions related to the features seen in bulk properties. Spectra were taken every five or ten degrees from -190 to +120 °C, and a selection is shown in Figure 4a. As is common in these systems the main features are broad disorder-coupled density-of-modes bands that persist even into the cubic paraelectric phase. Nonetheless there is finer structure that shows signatures of phase transitions, and in these samples those are seen as weak features at ~ 68, ~ 120 and ~150 cm$^{-1}$. Two of these have been fitted with Lorentzian lines to provide the temperature-dependent amplitude data in Fig. 4b. A signature of the T– C phase transition is seen in the loss of the intensity of the vibrational line at 150 cm$^{-1}$ at 85 ºC. However, the peak is still seen very weakly at temperatures up to 100 ºC, suggesting that the transition to the paraelectric phase is incomplete, that there are regions retailing T symmetry into the predominantly C phase. Neither this nor any phase transition has shown any evidence of hysteresis in their Raman spectra.

The line at 68 cm$^{-1}$ in the lowest-temperature (-190 ºC) begins to broaden and weaken as the temperature is raised, then is lost into a strong central peak by -40 ºC. Such behaviour is well-documented in similar ferroelectric perovskites[14,15], supporting the existence of a ferroelectric-ferroelectric phase transition roughly above -40 ºC. The fitted amplitude in Fig. 4b shows an extrapolated amplitude falling to zero at a transition temperature of -60 ºC, but with a weaker remnant persisting to -50 ºC. Elastic loss shows a weak anomaly at this temperature, marked by a dot-dashed line (III → III-b) in Fig. 2.

An ambient-temperature structural phase transition (marked by III-a → II in Fig. 2) can be associated with the disappearance of the vibrational line at 120 cm$^{-1}$, which lies between the two labelled features in Fig. 4a. This weak line overlaps with other features in the spectra and reliable fits could not be obtained. However, it appears across the entire lower-temperature range, losing strength and finally disappearing into noise by 30 ºC. The Raman data thus signal two transitions, at -55±10 ºC and 90±10 ºC, and hint at another near 30 ºC.

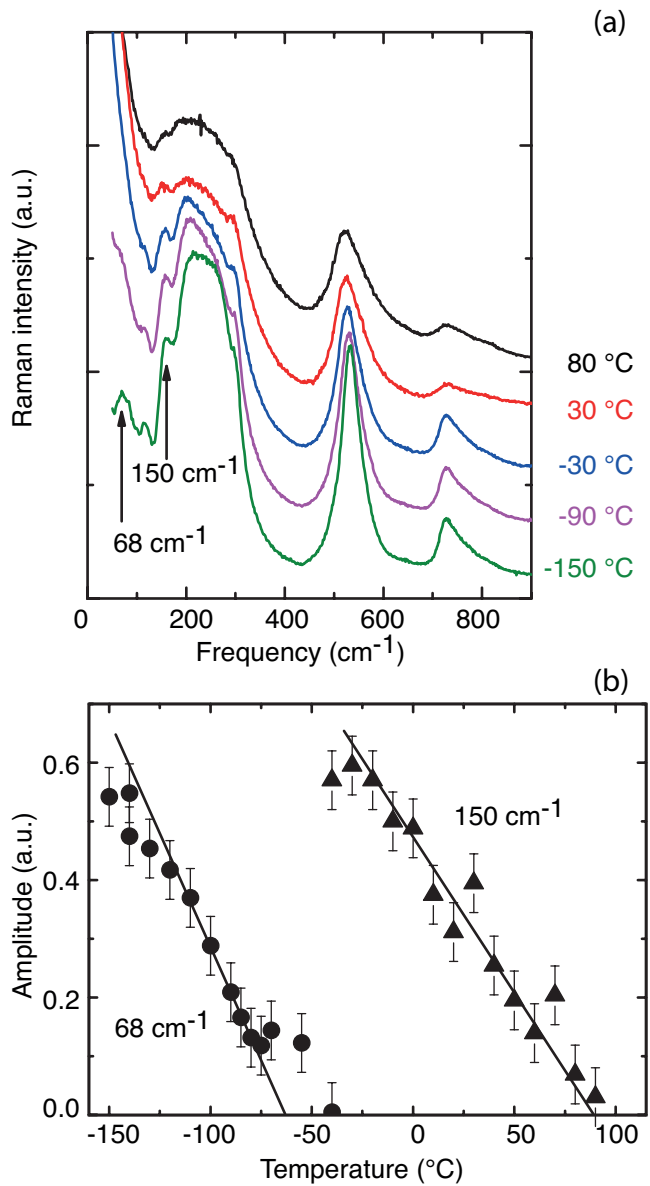

Figure 4 (color online). (a) Raman spectra at selected temperatures show a signature of a transition near 90 °C as a general reduction of structure, replacing it with simple broad density-of-modes bands, most clearly seen in the loss if intensity in the 150 cm$^{-1}$ feature (see also (b)). Note that these spectra have been normalised to their average intensity, which has artificially removed an approximately halving of intensity across the 90 °C transition. The line below 100 cm$^{-1}$ weakens and merges into a central peak near -60 °C, and the feature near 120 cm$^{-1}$ is lost near ambient temperature. (b) Fitted intensities for lines at 68 and 150 cm$^{-1}$.

In agreement with Raman data, evidence for a polar tetragonal phase above $T_C$ is confirmed by the piezoelectric signal which persists some 30°C above the peak in permittivity. Fig. 5 shows the height of conductance *G* at the resonant frequency of the planar vibration mode (∝ piezoelectric coefficient $d_{31}$, Ref. 16) and susceptance B of the sample (=ωC, where C is capacitance of the sample and ω frequency). Residual piezoelectricity detected in the paraelectric phase unequivocally proves that the macroscopic symmetry of the poled sample remains polar some 30°C above Curie temperature where centrosymmetric structure is expected. The probable origin of the polar symmetry are polar regions which survive the transition temperature within nonpolar matrix; this could obviously be caused by the chemical inhomogeneities or presence of defects in the material, as shown for $BaTiO_3$.[17,18] It is interesting that the temperature region where piezoelectric signal disappears (~120°C) corresponds to the onset of softening of the sample on cooling (compare Fig. 3 and Fig. 5). This is expected because electro-mechanical coupling leads to softening of a piezoelectric material.[19]

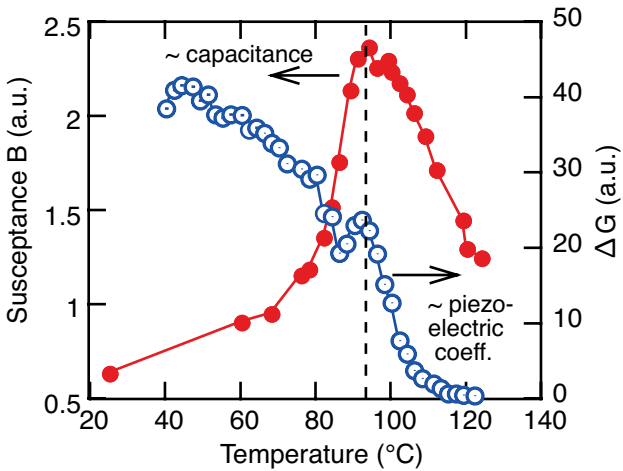

Figure 5 (color online). Susceptance B measured at 100 kHz and height of conductance peak, ΔG = G($ω_{resonance}$)- G(background). The resonant frequency of the sample was about 360 kHz at room temperature. The vertical dashed line marks the cubic-tetragonal transition temperature (II → I, see Fig. 2).

We now turn to the question of the origin of elastic and dielectric anomalies referring to regions I, II, III-a, III-b and III in Fig. 2. There is no doubt that I→II transition around 90°C is the same C-T phase transition as in $BaTiO_3$. As indicated by XRD data in Ref. 2 and our own, the second anomaly (II→III-a) marks a transition into a region consisting of a mixture of T and another phase identified in Ref. 2 as an "R" phase. The pure R phase in BCZT-50/30/20 is identified in Refs. 2 and 8 at -63°C and -180°C. Assuming that this assignment is correct, we can then identify phase transition detected in Raman spectra at -60°C (III→III-b) as a transition from that same R phase. The nature of phases in region III-b and "R" phase in III-a is less clear and several possibilities can be considered; we propose the following.

A complex sequence of several phase transitions is not rare in ferroelectrics: they may signal, for example, presence of incommensurate and re-entrant phases, such as reported in $Ba_2NaNb_5O_{15}$.[20,21] Thus, it is not impossible that the R phase from region III (below -60°C) may reappear in region III-a, mixed with the T phase. An additional possibility is that phases in regions III-a and III-b are rhombohedral or lower symmetry phases[22] different from each other and from the one below -60°C. A well-known example of such transitions is found in PZT, where high temperature R-phase with symmetry R3m transforms into a low-temperature rhombohedral R3c phase.[22-24] The difference between the two phases are oxygen octahedra tilts in the latter, which are seen in XRD as superlattice reflections; these may not be visible at the high angles shown in Ref. 2. Moreover, it would be important to see if the XRD signal of the mixed phase region III-a, which contains nanodomains,[8] can be explained with a mixture of T and O or T and monoclinic (M) phases. Finally, the anomaly immediately below room temperature may indicate a domain-wall relaxation process such as observed in $BaTiO_3$[25] and PZT.[26] However, being thermally activated, domain-wall processes are usually frequency dependent.[25,26] In the case of BCZT-50/30/20, the dielectric peaks were observed over four orders of magnitude (100 Hz- 1 MHz) whereas elastic anomaly was measured at 1 Hz by DMA and at 360 kHz by the resonant technique. The three anomalies (I→II, II→III-a, and III-a→III-b) show a comparably weak and irregular dependence on frequency, suggesting against a thermally activated mechanism.

In summary, we report Raman spectra and temperature-induced anomalies in dielectric and elastic coefficients of BCZT-50/30/20. The anomaly just below room temperature (III-a→III-b) could indicate a structural phase transition in addition to two reported earlier in the literature.[2] Because this anomaly appears near the operating (ambient) temperature it is likely that it contributes to the large electro-mechanical response observed in this material, either intrinsically or through its effect on domain structure. The present study cannot entirely eliminate the possibility that this anomaly originates from relaxation processes associated with interaction of domain walls with defects. However, such possibility does not appear to be consistent with the absence of a significant frequency dependence of the temperature at which anomaly appears. Another phase transition, which leaves a weak signature in $E''$, is signaled by Raman spectra at about -60°C,

The authors acknowledge financial support of FNS PNR62 project No. 406240-126091.

Note: after this manuscript was submitted for publication we became aware of the paper by F. Benabdallah, A. Simon, H. Khemakhem, C. Elissalde, and M. Maglione, "Linking large piezoelectric coefficients to highly flexible polarization of lead free BaTiO3-CaTiO3-BaZrO3 ceramics," *J. Appl. Phys.*, vol. 109, p. 124116, 2011. We suggest that the peak in the pyroelectric current observed below room temperature by Benabdallah et al. has the same origin as the elastic anomaly reported in our paper; i.e., the peak may signal a phase transition.